# High-Throughput In-Situ Fabrication of Fibrous Membranes Enables Scalable Passive Radiative Cooling

Hanzhuo Shao,[1] Xiaoli Huang,[2] Xuemei Huang,[3] Jin Zhao,[3] Nailin Xing,[4] Hua Xu,[2,*] Weijie Song,[3,†] and Yuehui Lu,[2,‡]

[1]College of Chemical Engineering, Zhejiang University of Technology, Hangzhou 310014, China

[2]School of Physical Science and Technology, Ningbo University, Ningbo 315211, China

[3]Ningbo Institute of Materials Technology and Engineering, Chinese Academy of Sciences, Ningbo 315201, China

[4]Institute of Vegetables, Ningbo Academy of Agricultural Sciences, Ningbo 315040, China

---

[*] Corresponding author: xuhua@nbu.edu.cn

[†] Corresponding author: weijiesong@nimte.ac.cn

[‡] Corresponding author: luyuehui@nbu.edu.cn

**Abstract**

Deploying fibrous membranes for passive daytime radiative cooling (PDRC) on large and irregular surfaces is highly desirable but remains challenging, owing to the slow deposition rates and the need for electrically conductive substrates in conventional electrospinning. Here, we demonstrate a high-throughput in-situ strategy for fabricating nanocomposite PDRC fibrous membranes via solution blow spinning. This method achieves deposition rates 8-12 times faster than electrospinning and can be applied directly onto nonplanar, nonconductive objects. The resulting membranes, composed of styrene-ethylene-butylene-styrene (SEBS) fibers embedded with $Y_2O_3$ nanoparticles, achieve sub-ambient cooling of up to 7.0 °C outdoors, effectively delaying ice melting. Moreover, they are fully recyclable through simple cleaning, dissolution, and reprocessing. This scalable and sustainable fabrication route provides a versatile and practical platform for integrating PDRC fibrous membranes across diverse surfaces, paving the way toward real-world thermal management applications.

**Keywords**

Passive daytime radiative cooling; fibrous membranes; high-throughput in-situ fabrication; solution blow spinning; recyclability.

# 1. Introduction

With the intensifying challenges of global warming and rising energy demands, the development of high-efficiency and environmentally sustainable cooling technologies has become increasingly urgent [1]. Among them, passive daytime radiative cooling (PDRC) has emerged as a promising solution, owing to its ability to provide cooling without electricity consumption [2,3]. PDRC lies on materials that combine high solar reflectance (0.3–2.5 μm) to suppress solar heating with strong infrared emissivity in the atmospheric window (8–13 μm) to radiate heat efficiently into outer space [4,5]. Such spectral features enable PDRC materials to achieve sub-ambient cooling under direct sunlight in an energy-free and environmentally benign manner [6,7].

In recent years, diverse PDRC materials have been reported, including multilayer photonic structures [8,9], metamaterials [10], porous polymers [11], structural woods [12], fabrics [13,14], coatings [15–19], and thin films [20,21]. Among them, fibrous membranes are attractive for covering large and irregular surfaces, offering both performance and cost advantages. To date, electrospinning has been the dominant technique for producing PDRC membranes with high solar reflectance and mid-infrared emittance typically exceeding 0.9 [22–26]. But it suffers from extremely low deposition rates (1–20 mL/hr, requiring nearly 300 hours to fabricate a square meter of membrane) [27,28]. Moreover, electrospinning relies on expensive high-voltage power supplies (typically tens of kilovolts) and precise nozzle systems, posing challenges for high-throughput fabrication and large-scale deployment [29,30]. Alternative fiber-producing approaches have also been explored: meltblowing achieves faster deposition rates (50–150 mL/hr) but yields coarse fibers with poor uniformity [31–33]; centrifugal spinning can produce more uniform fibers, yet still operates at limited rates (20–50 mL/hr) and requires complex equipment [34]. By contrast, solution blow spinning technology, first introduced in 2009 [35], has been widely applied in textiles [36], electronics [37], biomedical [38,39], and membrane separation [40]. Solution blow spinning combines potential high deposition rates with compatibility for nonplanar and non-conductive substrates, while avoiding the need for sophisticated equipment or high-voltage power supplies [38]. Very recently, solution blow spinning has been employed to fabricate PDRC materials, verifying the feasibility of this approach while being constrained by relatively low deposition rates of 25–30 mL/hr [41,42].

In this work, we demonstrate solution blow spinning as a high-throughput strategy for fabricating PDRC fibrous membranes. By optimizing parameters such as solution concentration, viscosity, air pressure, and working distance, we fabricated nanocomposite membranes composed of styrene-ethylene-butylene-styrene (SEBS) fibers embedded with $Y_2O_3$ nanoparticles. We achieved an average deposition rate remained around 200-300 mL/hr, reaching the deposition rates 8-12 times faster than conventional electrospinning [29,35,38]. The membranes achieve sub-ambient cooling of up to 7.0 °C outdoors, effectively delaying ice melting. They can be deposited in situ onto nonplanar and non-conductive substrates, including agate, ceramic, and glass containers, and can also be delaminated through tailoring their amphiphobic wettability. Moreover, the membranes exhibit excellent mechanical flexibility and are fully recyclable through rapid cleaning, dissolution, and reprocessing within half an hour. Importantly, this solution blow spinning-based strategy is not limited to SEBS-$Y_2O_3$ nanocomposites but is readily extendable to other polymer-solvent systems and even inorganic fibrous membranes. Taken together, our work establishes a universal and scalable platform for sustainable fabrication of PDRC fibrous membranes across diverse surfaces, paving the way for real-world thermal management applications.

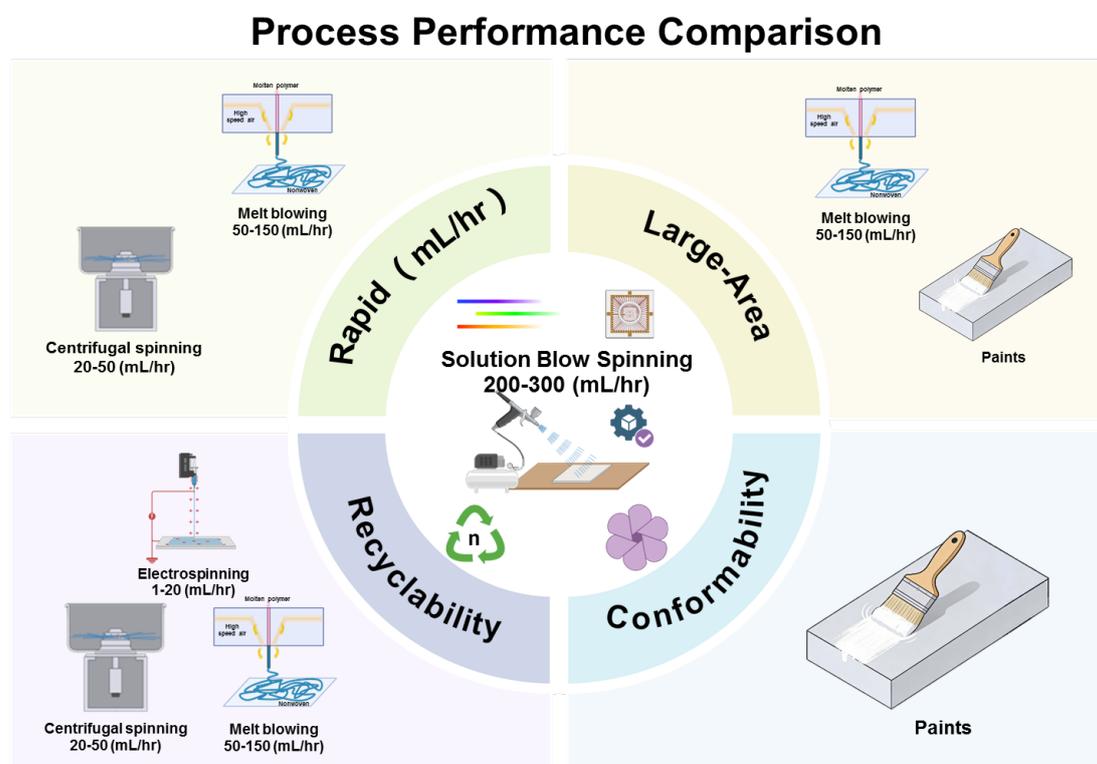

**Figure 1.** Performance comparison of various fabrication processes with respect to production speed, large-area scalability, conformability to complex surfaces, and recyclability. SEBS, PLA (polylactic acid), and cellulose are selected as respective materials for benchmarking.

## 2. Experiment and characterization

### 2.1. Materials

The $Y_2O_3$ powders used in this study were purchased from Suzhou Youzirconium Nanomaterial Co., Ltd. The SEBS block copolymer (G1645) was supplied by Kraton Corporation. Cyclohexane (Analytical Reagent, 99.5%) was obtained from Aladdin Biochemical Technology Co., Ltd. All chemicals were used as received without further purification.

## 2.2. Preparation of PDRC Fibrous Membranes

SEBS and $Y_2O_3$ powders were blended at a predetermined mass ratio and dissolved in cyclohexane under magnetic stirring for two hours to yield a homogeneous solution. The resulting solution was loaded into an airbrush nozzle, where compressed gas (0.1–3 bar) propelled the jet onto target substrates at a high deposition rate of 200-300 mL/hr. The SEBS-$Y_2O_3$ nanocomposite fibrous membranes were directly formed in situ on diverse substrates at controlled ambient conditions (25 ± 2 °C, 50 ± 5% RH) in a standard laboratory environment. Rapid solvent evaporation during jet flight enabled in situ formation of dry membranes, without additional post-drying.

## 2.3. Characterizations

The dynamic viscosity of the polymer solution was measured using a viscometer (NDJ-8S, Sunne) with a No. 2 spindle at 60 rpm. The surface morphology of the fibrous membranes was characterized by a field-emission scanning electron microscope (FE-SEM, Regulus 8230, Hitachi). The solar spectral reflectance (200–2500 nm) was measured with a ultraviolet-visible-near infrared (UV-Vis-NIR) spectrophotometer (Lambda 1050, PerkinElmer). Infrared transmittance ($T$) and reflectance ($R$) spectra over the range of 3–20 μm were recorded using a Fourier-transform infrared (FTIR) spectrometer equipped with a gold-coated integrating sphere (Nicolet 6700, Thermo Scientific). The spectral emissivity ($\varepsilon$) was determined according to $\varepsilon = 1 - R - T$. The thermal conductivity of the membranes was measured using a thermal constant analyzer (TPS3500, Hot Disk). The mechanical properties of the membranes were evaluated through tensile testing on a universal testing system (Instron 3365) using samples with a thickness of ~0.46 mm and a width of 1.2 mm. The UV aging resistance of the membranes was tested using a UV aging chamber (UMC1200, Shanghai Yixiang Test Equipment Co., Ltd.).

During outdoor testing, the ambient temperature was monitored with a data logger (RDXL4SD, Omega Engineering), while relative humidity and solar irradiance were measured using a pyranometer (PTS-3, Sunshine Weather Technology). Wind speed was recorded with an anemometer (Kestrel NK5500). Water and oil (diiodomethane) contact angles were measured using an optical contact angle system (CA 200, Guangdong Beidou Precision Instrument Co., Ltd.).

## 3. Results and Discussion

The Ohnesorge number ($Oh$) is a key dimensionless parameter that characterizes the spinnability of a solution [29]; values much greater than unity generally indicate favorable conditions for fiber formation [43]. According to Eq. (1), which relates the Weber ($We$) and the Reynolds ($Re$) numbers, the Ohnesorge number is expressed as:

$$Oh = \frac{\sqrt{We}}{Re}, \tag{1}$$

where the Weber and the Reynolds numbers can be calculated to be $We$ = 0.238 and $Re$ = 0.253. The calculation yields $Oh$ = 1.93 (see Supporting Information, Note S1). This calculation is based on a viscosity ($\mu$) of 46.54 mPa·s (Fig. S1), a density ($\rho$) of 0.92 g/cm³ (estimated from the solution mass and volume), a surface tension ($\sigma$) of 25.0 mN/m, a nozzle diameter ($D$) of 0.8 mm, and an effective diameter ($d$) of 25.3 μm, assuming that the jet is divided into 1000 streams. The resulting high Ohnesorge number confirms that the solution lies within the spinnable regime, ensuring reliable fabrication of fibrous membranes, as demonstrated in Fig. 2a [29,43].

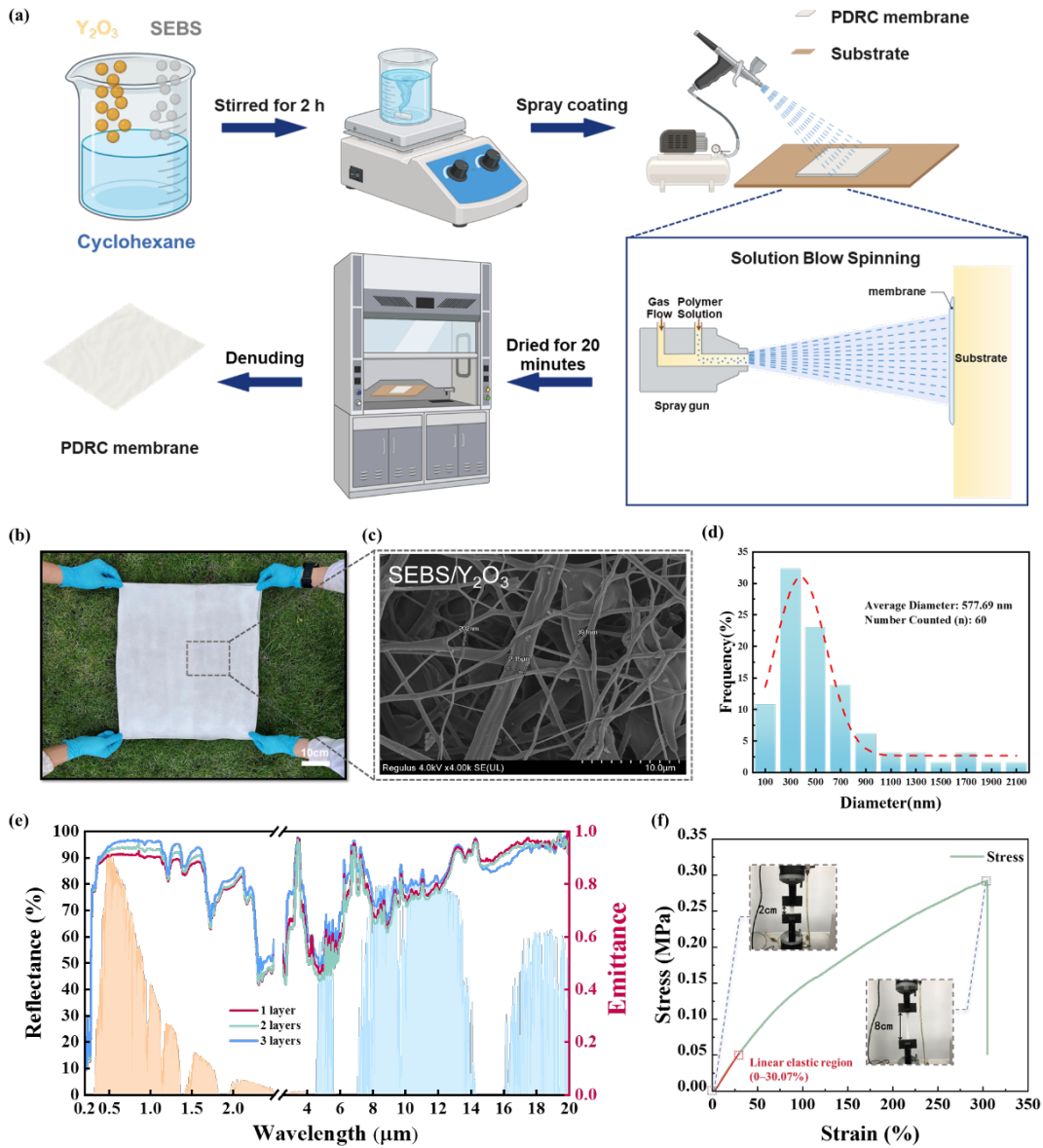

**Figure 2. Fabrication and characterization of the SEBS-Y$_2$O$_3$ nanocomposite fibrous membrane.** (a) Schematic illustration of the high-throughput in-situ fabrication process via solution blow spinning. (b) Photograph of a large-area, free-standing membrane. (c) SEM image of SEBS-Y$_2$O$_3$ nanocomposite membrane. (d) Diameter distribution of the fibers in the membrane. (e) Spectral reflectance and emissivity of membranes with different layer numbers and thicknesses. (f) Stress-strain curve of the membrane, showing an elongation at break exceeding 300%, tensile strength of 0.29 MPa, fracture load of 1.61 N, and an elastic modulus is 0.18 MPa determined from the linear elastic region within 0–30.07% strain.

A SEBS-Y$_2$O$_3$ nanocomposite fibrous membrane with an area of approximately 500×500 mm$^2$ was fabricated via solution blow spinning at an ultrahigh feed rate of 200-300 mL/hr (Fig. 2b and a larger-scale membrane in Fig. S2). Based on the measured areal mass and raw material prices, the estimated material cost is approximately 5.0 USD m$^{-2}$ (see Supplementary Information, Note S2), indicating the economic feasibility for large-area applications. SEM image (Fig. 2c) reveals a highly porous morphology with a broad fiber-diameter distribution. The single-layer cross-sectional morphology (Fig. S3) further confirms the interconnected fibrous network. Statistical analysis based on 60 fibers (Fig. 2d) shows that the diameters span from ~100 to > 2000 nm, with an average diameter of ~578 nm. The distribution exhibits a dominant peak in the 300–500 nm range, followed by a rapid decay at larger diameters, indicating a broad yet statistically stable size profile. Such pronounced diameter heterogeneity forms a porous fibrous network, which is advantageous for simultaneously enhancing solar reflectance and mechanical robustness. As confirmed by the scattering efficiency simulations of SEBS fibers with varying diameters and wavelengths (Fig. S5), the fibers exhibit strong light-scattering capability at ~578 nm. Furthermore, the Y$_2$O$_3$ particle size used in this work lies within the reported optimal range of 0.5–6.5 μm for PDRC polymer composites (including SEBS) [44]. The TEM image and corresponding EDS mapping (Fig. S4) confirm that the Y$_2$O$_3$ nanoparticles are uniformly distributed within the SEBS fiber matrix without noticeable agglomeration, providing direct microscopic evidence for the formation of SEBS-Y$_2$O$_3$ composite fibers. Optically, Y$_2$O$_3$, with a wide bandgap of 5.6 eV and a high refractive index exceeding 1.9 [44], is beneficial for enhancing light-matter interaction and scattering in the solar spectral region. Such characteristics, combined with the fibrous architecture, are conducive to the synergistic optical behavior of the composite membrane relevant to PDRC applications.

To comprehensively evaluate the PDRC performance of the SEBS-$Y_2O_3$ membranes, their optical properties were systematically characterized, with a focus on solar reflectance and infrared emissivity. The solar-band reflectance spectra of SEBS-$Y_2O_3$ membranes with different layer numbers are shown in Fig. 2e. A single SEBS-$Y_2O_3$ layer (thickness of ~0.46 mm) exhibits an average reflectance of 88.93%, indicating effective suppression of incoming sunlight. Analysis of the single-layer spectra (Fig. S6) reveals moderate transmittance and negligible absorbance in the solar band. With increasing thickness, the transmittance decreases markedly while absorbance remains nearly constant, resulting in enhanced reflectance. Accordingly, the reflectance increases to 92.19% for two layers and 93.31% for three layers, demonstrating that multilayer stacking effectively improves the solar-reflective capability. The corresponding infrared emissivity spectra (Fig. 2e) show that the average emissivity is 0.76, 0.74, and 0.79 for one-, two, and three-layer membranes, respectively, indicating consistently high thermal emissivity across all configurations. This behavior is primarily attributed to the incorporation of $Y_2O_3$ nanoparticles as mid-infrared emissive fillers, which provide strong phonon-mediated radiation within the atmospheric window. Although slight fluctuations are observed with increasing layer number, the three-layer membrane maintains a high emissivity, ensuring effective thermal radiation to outer space. The synergetic combination of high solar reflectance and high infrared emissivity is essential for efficient PDRC. The former minimizes solar heat absorption, while the latter promotes rapid heat dissipation, benefiting sub-ambient cooling.

Superior mechanical properties are essential for the practical application of PDRC membranes, which may experience various mechanical stresses in real-world environments, such as wind loads and compressive forces. The mechanical performance of the SEBS-$Y_2O_3$ membranes was evaluated by tensile testing. As shown in the stress-strain curve (Fig. 2f), the membrane exhibits an elongation at break of 304.72%, together with a tensile strength of 0.29 MPa (stress at break), fracture load of 1.61 N, and elastic modulus of 0.18 MPa, determined from the slope of the linear elastic region within 0–30.07% strain. These metrics collectively reflect the membrane's excellent flexibility, ductility, tensile resistance and deformation resistance, supporting its reliability against outdoor-relevant mechanical stresses. During the test, the SEBS-$Y_2O_3$ membrane was stretched from an initial length of 2 cm to over 8 cm prior to failure, confirming its high mechanical stability under large deformation. Such outstanding stretchability enables the SEBS-$Y_2O_3$ membranes to conform to complex geometries and dynamic mechanical loading while maintaining structural integrity, which is crucial for long-term durability in practical applications.

**3.1. Outdoor Temperature Measurements**

Outdoor temperature measurements were conducted to evaluate the PDRC performance of the SEBS-$Y_2O_3$ membranes under natural conditions. The experimental setup and measuring instruments are shown in Fig. 3a. As schematically illustrated, a 12 × 12 × 6 $cm^3$ acrylic box was used as the test chamber. To minimize the influence of solar irradiation on the internal temperature, the outer surface of the chamber was covered with aluminum foil [24,45–49]. In addition, the chamber interior was lined with low-thermal-conductivity polystyrene foam to suppress heat exchange with the environment. During the measurements, the temperatures of the SEBS-Y$_2$O$_3$ membranes was monitored in real time using thermocouple probes attached to the backside of the membranes. The measurements were conducted on a sunny day, and meteorological parameters, including ambient temperature, relative humidity, solar irradiance, and wind speed (Fig. S7), were recorded throughout the test. The variations in relative humidity and solar irradiance are shown in Fig. 3b. The peak solar irradiance reached ~1090 W/m$^2$ during the test period. Although fluctuations in relative humidity and solar irradiance influenced the cooling behavior to some extent, the SEBS-Y$_2$O$_3$ membranes consistently exhibited stable and efficient cooling performance.

To quantitatively evaluate the PDRC performance of the SEBS-Y$_2$O$_3$ membranes, a cooling power measurement was conducted over the same period as the outdoor temperature testing. During the experiment, the backside of the SEBS-Y$_2$O$_3$ membranes was thermally coupled to an electric heater through a 1-mm-thick copper plate, utilizing the high thermal conductivity of coper to ensure efficient heat transfer. A feedback control system was employed to maintain the membranes at ambient temperature. During the measurement, both the ambient and membrane temperatures were continuously monitored by the feedback system, which dynamically adjusting the input voltage of the heater to maintain thermal equilibrium. This approach effectively eliminated the influence of ambient temperature fluctuations, enabling accurate determination of the net cooling power. As shown in Fig. 3c, the SEBS-Y$_2$O$_3$ membranes exhibited stable cooling performance throughout the test period, with an average cooling power of 23.1 W/m$^2$. In addition to the experimental measurement, the net cooling power was also theoretically evaluated under the standard AM1.5G solar irradiance of 1000 W/m² (see Supplementary Information, Note S3 and Fig. S8 for details). The calculation shows a net cooling power of 18.8 W/m² and a sub-ambient cooling of 2.0 °C. These values are in close agreement with the experimental results obtained under comparable environmental conditions.

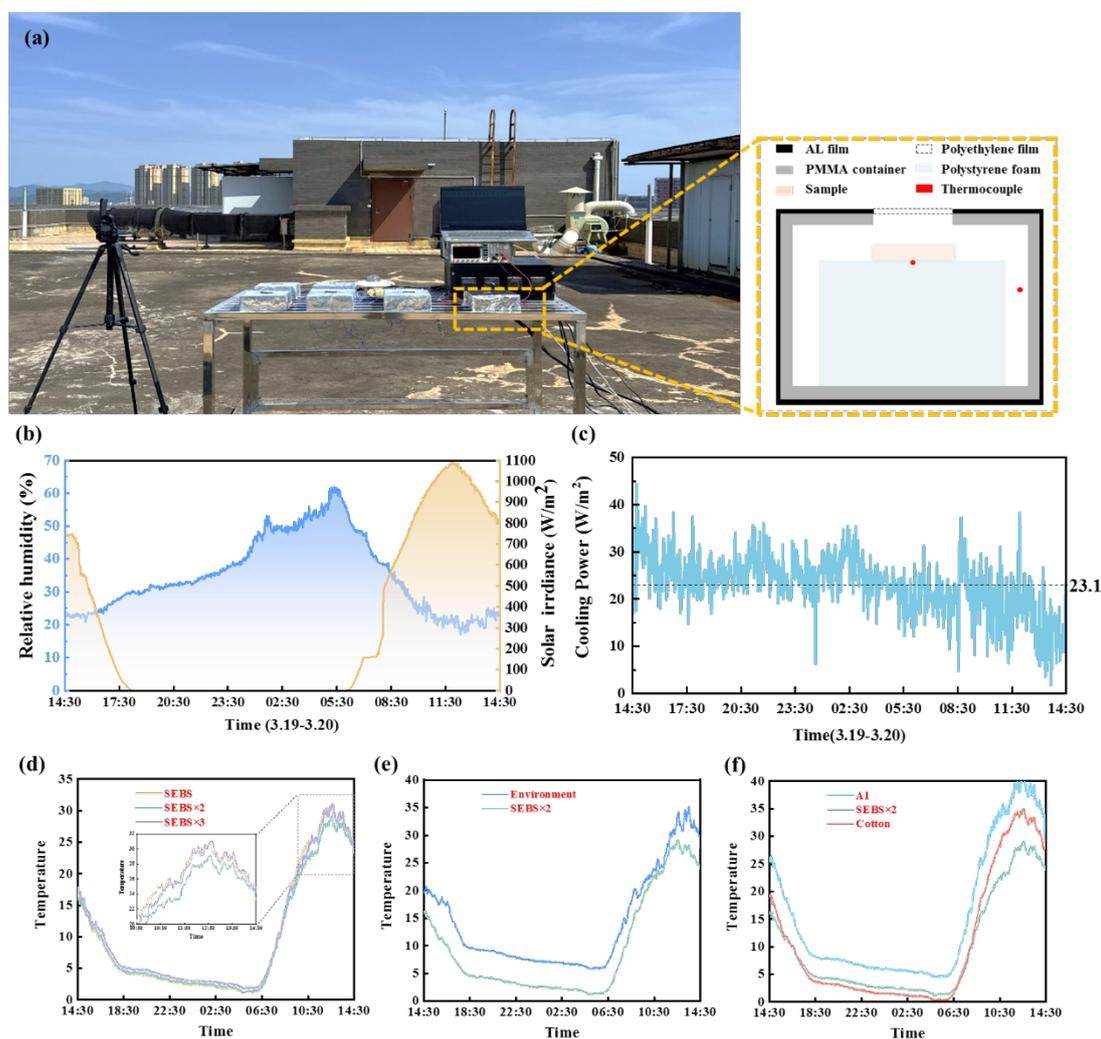

**Figure 3. Outdoor cooling performance evaluation of the SEBS-Y₂O₃ membranes.** (a) Photograph of the outdoor testing setup (left) and a schematic of the custom acrylic chamber designed to minimize environmental heat transfer effects (right). (b) Recorded environmental parameters during the test: relative humidity and solar irradiance. (c) Measured cooling power of the SEBS-Y$_2$O$_3$ membrane over the testing period. (d) Temperature profiles of SEBS-Y$_2$O$_3$ membranes with different layer numbers and thicknesses, plotted together with the ambient temperature. (e) Sub-ambient temperature difference of the two-layer SEBS-Y$_2$O$_3$ membrane, highlighting its continuous cooling capability during both daytime and nighttime. (f) Temperature comparison among the two-layer SEBS membrane, a cotton cloth, and an aluminum foil control sample under direct sunlight.

During the outdoor temperature test, the temporal temperature evolution of the membranes was continuously recorded. The studied samples included the SEBS-Y$_2$O$_3$ membranes with 1–3 layers (single-layer thickness of 0.46 mm), with cotton cloth and aluminum foil serving as control samples. Cotton cloth was selected as a baseline material due to its similar fibrous morphology, comparable optical characteristics (low solar absorption and relatively high emissivity), and close thermal conductivity (0.06 W m$^{-1}$ K$^{-1}$ for the cotton) to the SEBS-Y$_2$O$_3$ membrane (Fig. S9 and 0.08 W m$^{-1}$ K$^{-1}$). Aluminum foil was adopted as a second control because it possesses high solar reflectance but very low mid-infrared emissivity. This contrast allows for qualitative evaluation of cooling contribution beyond solar reflection. As shown in Fig. 3d, comparison of the temperature evolution beneath SEBS-Y$_2$O$_3$ membranes of different thicknesses with those beneath the control samples provide a direct assessment of the cooling performance of the SEBS-Y$_2$O$_3$ membranes. Under these conditions, the SEBS-Y$_2$O$_3$ membranes exhibited a pronounced cooling effect. Notably, the two-layer SEBS-Y$_2$O$_3$ membranes (total thickness of 0.92 mm) maintained a temperature below the ambient temperature from 14:30 on the test day to 14:30 on the following day, with a maximum sub-ambient temperature reduction of 7.0 °C, a minimum of 0.8 °C, and an average daily sub-ambient temperature reduction of 4.6 °C (see Fig. 3e). As shown in Fig. 3f, compared with the cotton-cloth control, the two-layer SEBS-$Y_2O_3$ membrane maintained a lower daytime temperature, with a maximum temperature reduction of 7.7 °C. Moreover, the two-layer SEBS-$Y_2O_3$ membrane consistently remained cooler than aluminum foil, with a maximum and minimum temperature reduction of 13.3 and 3.0 °C, respectively. These results demonstrate that SEBS-$Y_2O_3$ membranes can effectively lower the surface temperature and achieve pronounced cooling under natural outdoor conditions. The superior cooling performance is primarily attributed to the combination of high solar reflectance and high infrared emissivity due to the porous nanocomposite fibrous architecture of the membranes, which enables efficient suppression of solar heat gain and rapid thermal radiation to outer space.

These results confirm that the SEBS-$Y_2O_3$ membranes can deliver substantial cooling power under practical outdoor conditions. Through these quantitative cooling-power measurements, the efficiency of the SEBS-$Y_2O_3$ membranes is further validated, highlighting their strong potential for applications in energy-efficient buildings, electronic device cooling, and personal thermal management.

**3.2. High-Temperature Outdoor Ice-Preservation Experiment**

To intuitively validate the practical cooling efficacy of the SEBS-$Y_2O_3$ membranes, a comparative experiment was performed. Figure 4a presents a schematic illustration of the experimental setup and the comparative samples. Four ice samples (13.4 g each) were prepared with different wrappings: (1) SEBS-$Y_2O_3$ membrane, (2) cotton cloth, (3) aluminum foil, and (4) an unwrapped control. The experiment was conducted in an outdoor environment with an average temperature of 38 °C, and the melting time was precisely recorded. The corresponding ambient conditions during the test are shown in Fig. 4b. During the experiment, photographs were taken at each time interval (Fig. S10), together with images of the ice blocks during weighing, to ensure the accuracy and reproducibility of the measurements. In addition, continuous temperature records were collected to track variations in the ambient conditions for confirming the stability and effectiveness of the SEBS-Y$_2$O$_3$ membranes under high-temperature outdoor environments.

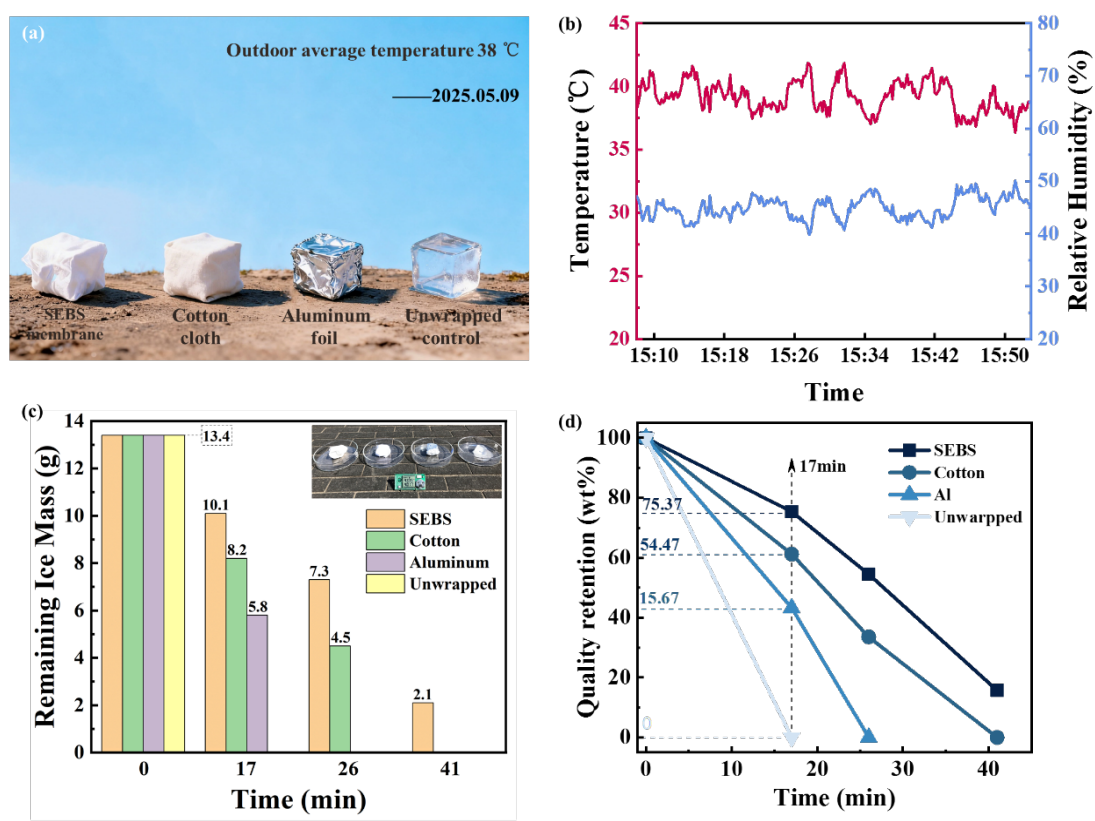

**Figure 4. Ice preservation capability of the SEBS-Y$_2$O$_3$ membranes.** (a) Schematic of the outdoor ice-melting test setup. (b) Recorded ambient temperature and relative humidity during the test. (c) Remaining ice mass as a function of time with different wrapping configurations, and (d) corresponding mass-retention percentage for evaluation of ice-preservation performance.

As summarized in the bar charts in Figs. 4c and 4d, the four samples exhibit distinct ice-preservation performance. The unwrapped control ice completely melted within 17 min, followed by the aluminum foil-wrapped ice at 26 min and the cotton-wrapped ice at 41 min. In sharp contrast, the SEBS-$Y_2O_3$ membrane-wrapped sample retained 2.1g of solid ice at the endpoint of 41 min, demonstrating the lowest melting rate. The preservation time achieved by the SEBS-$Y_2O_3$ membrane is approximately 2.53 times longer than the unwrapped control, highlighting its superior cooling performance and strong capability for suppressing ice melting. This ice-preservation test provides direct experimental evidence of the excellent practical cooling performance of the SEBS films. Owing to their energy-free cooling characteristic, the SEBS-$Y_2O_3$ membranes hold significant promise for applications requiring low-temperature maintenance, including cold-chain transportation, food preservation, and temporary storage of medical supplies [50–52], particularly in outdoor or off-grid environments where a stable power supply is unavailable.

**3.3. Recyclability and environmental Sustainability**

The solution blow spinning process enables in-situ fabrication of SEBS-$Y_2O_3$ nanocomposite fibrous membranes on a wide variety of substrates with diverse shapes and compositions. As shown in Fig. 5a, uniform SEBS-$Y_2O_3$ membranes can be directly formed on agate, ceramics, glass, and 120-grit sandpaper, demonstrating excellent conformability and adaptability even to rough and irregular surfaces. This broad substrate compatibility not only enhances the versatility of the fabrication process but also facilitates the deployment of SEBS-$Y_2O_3$ membranes in a wide range of practical applications.

On the other hand, as visually confirmed in Fig. 5a, the SEBS-$Y_2O_3$ membranes can also be easily and cleanly demolded from the substrates, demonstrating excellent deformability and strong application potential. This outstanding demoldability originates from the intrinsic amphiphobicity of the SEBS-$Y_2O_3$ membranes, which is quantitatively verified in Figs. 5b and 5c. The measured contact angles for water and oil (diiodomethane) reached 124.5° and 118.8°, respectively. Such high contact angles indicate an extremely low surface energy, resulting in weak intermolecular interactions and significantly reduced adhesion between the membranes and the substrates [53,54]. Consequently, the energy barrier for demolding is significantly lowered. This facile demolding behavior is highly desirable in practical applications, as it not only improves production efficiency but also prevents damage to expensive or delicate substrates during removal. Moreover, the amphiphobic nature of the membranes endows them with strong resistance to contamination. Owing to the weak interactions between the membrane surface and environmental contaminants such as water and oil, the membranes exhibit excellent anti-fouling capability. An environmental stability evaluation was conducted via UV aging tests. The membranes were continuously exposed to a UV aging chamber (145.4 μW/cm$^2$ at 365 nm) for one month. After exposure, no noticeable yellowing or morphological degradation was observed (Fig. S11). The optical spectra reveal an average decrease of only ~2% in UV-band reflectance, while mid-infrared emittance remains nearly unchanged, demonstrating robust optical stability under intensive UV exposure. Together with the intrinsic amphiphobic surface characteristics, these results demonstrate that the membranes possess strong anti-fouling and UV-resistant capability, which is essential for maintaining long-term optical performance and reducing maintenance requirements in realistic outdoor applications.

In the development of PDRC membranes, sustainability is a critical factor determining their practical application potential. The SEBS-$Y_2O_3$ membranes exhibit remarkable recyclability [24,55–57], which enables them to be reprocessed through simple post-treatment after their service life. This recyclability effectively reduces material waste, lowers production costs, and significantly enhances overall sustainability. The recycling procedure for the SEBS-$Y_2O_3$ membranes is illustrated in Fig. 5d. Briefly, the aged and the demolded SEBS-$Y_2O_3$ membranes were first subjected to a simple cleaning process. Tap water was used for preliminary rinsing to remove surface dust and loosely attached impurities, followed by ultrasonic cleaning in deionized water to further eliminate residual contaminants. After ultrasonic treatment, the membranes were blow-dried to ensure complete surface dryness. The dried membranes were then crushed into small pieces, dissolved in cyclohexane under heating, and reconstituted into a spinning solution for reprocessing into recycled membranes via the same solution blow spinning. Compared to pristine membranes of identical thickness, the recycled membranes exhibited a slightly reduced solar reflectance but an enhanced mid-infrared emissivity (Fig. S12). This variation is likely attributed to trace residual impurities introduced during the cleaning process, which may subtly modify the surface roughness and thereby affect the optical response. Nevertheless, the recycling process remains highly efficient and practical: the entire procedure, from cleaning to reprocessing, can be completed within 30 minutes, which is significantly faster than the fabrication of new membranes from raw materials.

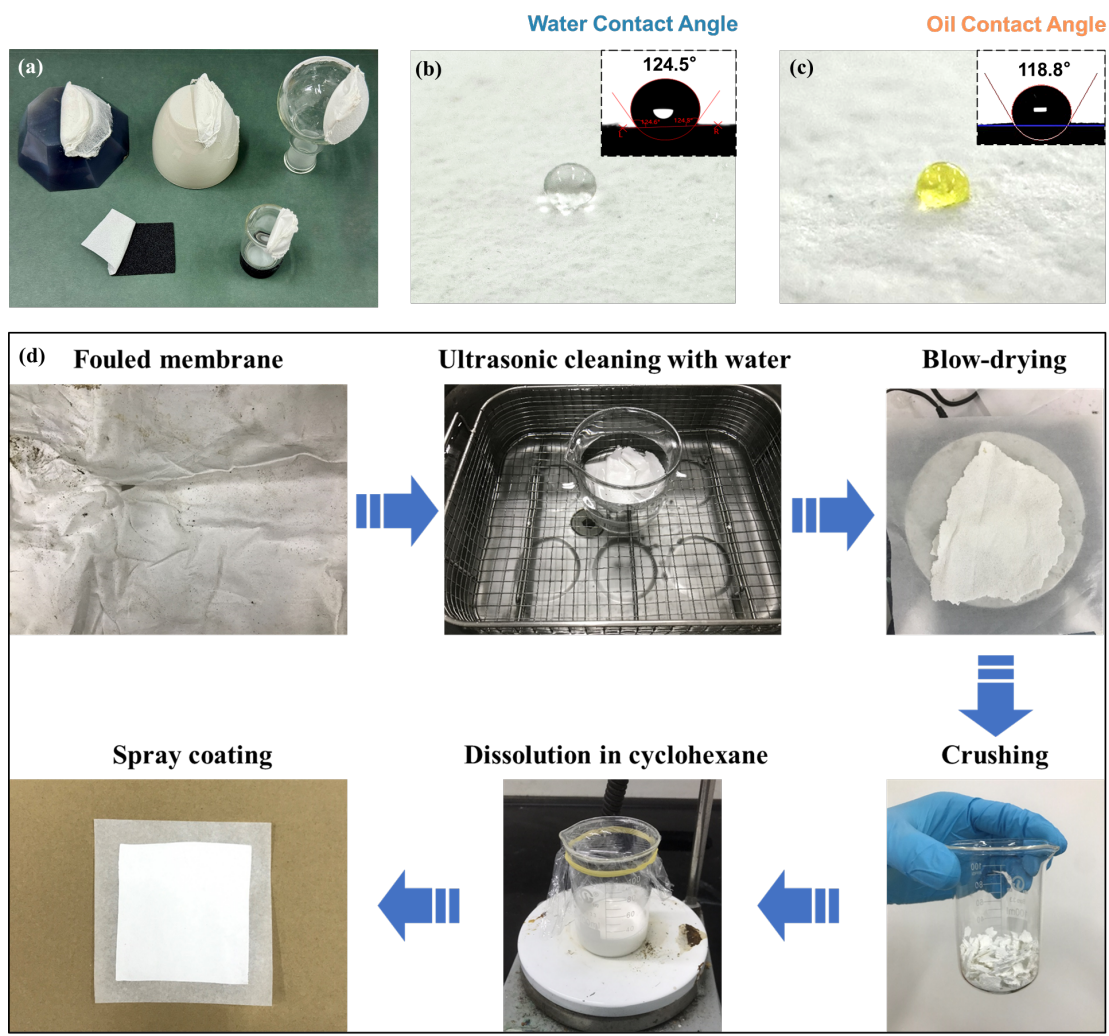

**Figure 5. Conformability and recyclability of the SEBS-Y$_2$O$_3$ fibrous membranes.** (a) Photographs demonstrating the conformability to various complex surfaces and the facile demoldability. (b and c) Optical images and corresponding measurements of the contact angles for water (124.5°) and (c) oil (diiodomethane, 118.8°), respectively, confirming the membrane's excellent amphiphobicity. (d) Schematic illustrating the simple and efficient recycling procedure of the SEBS-Y$_2$O$_3$ membranes.

This efficient recycling strategy significantly reduces material waste and production cost while markedly enhancing sustainability. Owing to the intrinsic hydrophobic and oleophobic nature of SEBS, the membranes can be cleanly demolded and remain mechanically self-supporting [58], which is essential for their facile recyclability. After simple cleaning and reprocessing, aged membranes can be directly reused, enabling a closed-loop material cycle. This recyclable and self-supporting feature not only complies with sustainable development principles but also substantially strengthens the practical feasibility and scalability of PDRC technologies.

## 4. Conclusions

In conclusion, we demonstrate a scalable and sustainable route for fabricating PDRC fibrous membranes via solution blow spinning. This high-throughput technique enables in-situ deposition of SEBS-$Y_2O_3$ nanocomposite membranes on large-area and nonplanar substrates at a production rate of 200-300 mL/hr, which is more than an order of magnitude higher than that of conventional electrospinning, addressing a key manufacturing limitation of fibrous PDRC materials.

The resulting membranes exhibit a favorable combination of optical and mechanical properties, including high solar reflectance (92.19%), mid-infrared emissivity (0.74), and excellent mechanical flexibility with a fracture strain exceeding 300%. Under outdoor conditions, the membranes achieve continuous sub-ambient cooling during both daytime and nighttime, with an average temperature reduction of 4.6 °C , at most 7.0 °C below ambient. In addition, the membranes effectively suppress ice melting at an ambient temperature of ~38 °C, extending the ice preservation time by a factor of ~2.53 compared with the unwrapped control.

Beyond cooling performance, sustainability is emphasized as a core design principle. Owing to their intrinsic amphiphobicity, the membranes can be cleanly demolded from diverse substrates and remain mechanically self-supporting. More importantly, they can be fully recycled through a rapid cleaning–dissolution–re-spinning process completed within 30 min, while largely retaining their optical cooling performance after recycling. This closed-loop recyclability significantly reduces material waste and production cost. Overall, this work establishes solution blow spinning as a practical and scalable platform for producing conformal, efficient, and recyclable PDRC membranes, paving the way for their deployment in energy-efficient buildings, cold-chain logistics, glacier protection, electronic cooling, and personal thermal management, particularly in outdoor and off-grid environments.

**Data availability**

The data that support the findings of this study are available from the corresponding author upon reasonable request.

**ACKNOWLEDGMENTS**

This work was supported by the National Natural Science Foundation of China (12174209), the Ningbo Key Laboratory of Silicon and Organic Thin Film Optoelectronic Technologies, and the Jiangsu Province Cultivation Base for State Key Laboratory of Photovoltaic Science and Technology.

**Competing interests**

The authors declare no competing interests.

**Additional information**

**Supplementary information** is available for this paper at https://doi.org/.

# Supporting Information:

# High-Throughput In-Situ Fabrication of Fibrous Membranes Enables Scalable Passive Radiative Cooling

*Hanzhuo Shao,[1] Xiaoli Huang,[2] Xuemei Huang,[3] Jin Zhao,[3] Nailin Xing,[4] Hua Xu,[2,*] Weijie Song,[3,†] and Yuehui Lu,[2,‡]*

[1]College of Chemical Engineering, Zhejiang University of Technology, Hangzhou 310014, China

[2]School of Physical Science and Technology, Ningbo University, Ningbo 315211, China

[3]Ningbo Institute of Materials Technology and Engineering, Chinese Academy of Sciences, Ningbo 315201, China

[4]Institute of Vegetables, Ningbo Academy of Agricultural Sciences, Ningbo 315040, China

---

[*] Corresponding author: xuhua@nbu.edu.cn

[†] Corresponding author: weijiesong@nimte.ac.cn

[‡] Corresponding author: luyuehui@nbu.edu.cn

**Note S1. Calculations of the Ohnesorge number**

**S1.1. Determination of the parameters in the Ohnesorge number**

**Nozzle diameter ($D$):**

$$D = 0.8\,\text{mm} = 8 \times 10^{-4}\,\text{m}$$

**Volumetric flow rate ($Q$):**

$$Q = \frac{15.27\,\text{ml}}{1\,\text{min}} = \frac{1.527 \times 10^{-5}\,\text{m}^3}{60\,\text{s}} \approx 2.545 \times 10^{-7}\,\text{m}^3/\text{s}$$

**Nozzle cross-sectional area ($A$):**

$$A = \pi \times \left(\frac{D}{2}\right)^2 = \pi \times \left(\frac{8 \times 10^{-4}}{2}\right)^2\,\text{m}^2 \approx 5.027 \times 10^{-7}\,\text{m}^2$$

**Fluid velocity ($V$):** The fluid velocity at the nozzle exit was calculated from the volumetric flow rate and the nozzle geometry:

$$V = \frac{Q}{A} = \frac{2.545 \times 10^{-7}\,\text{m}^3/\text{s}}{5.027 \times 10^{-7}\,\text{m}^2} \approx 0.506\,\text{m/s}$$

**Estimated number of filaments ($N$):** The flow is anticipated to divide into numerous fine filaments. An order-of-magnitude estimate of $N \approx 1000$ filaments was used for this initial calculation.

**Effective jet diameter ($d$):** The characteristic length of the jet is taken as the initial diameter of an individual fluid filament rather than the nozzle diameter. Assuming the flow splits into $N$ filaments with circular cross-sections, conservation of total cross-sectional area gives:

$$A = N \cdot A_{jet} \Rightarrow \frac{\pi D^2}{4} = N \cdot \frac{\pi d^2}{4} \Rightarrow d = \frac{D}{\sqrt{N}}$$

Substituting the values:

$$d = \frac{8 \times 10^{-4}\,\text{m}}{\sqrt{1000}} \approx 2.53 \times 10^{-5}\,\text{m}$$

**Dynamic viscosity (μ):** The viscosity was measured using a viscometer (NDJ-8S, Sunne), as shown in Figure S1:

$$\mu = 46.54\,\text{mPa·s} = 4.654 \times 10^{-2}\,\text{Pa·s}$$

**Density (ρ):**

$$\rho = \frac{14\,\text{g}}{15.27\,\text{ml}} = \frac{14 \times 10^{-3}\,\text{kg}}{15.27 \times 10^{-6}\,\text{m}^3} \approx 920\,\text{kg/m}^3$$

**Surface tension (σ):**

$$\sigma = 25\,\text{mN/m} = 2.5 \times 10^{-2}\,\text{N/m}$$

**S1.2. Calculated Reynolds number (Re):** The Reynolds number, which represents the ratio of inertial to viscous forces in the flow, is calculated as follows [1]:

$$Re = \frac{\rho V d}{\mu} = \frac{(920\,\text{kg/m}^3) \times (0.506\,\text{m/s}) \times (2.53 \times 10^{-5}\,\text{m})}{4.654 \times 10^{-2}\,\text{Pa·s}} \approx 0.253$$

A Reynolds number much small than unity ($Re \ll 1$) indicates that the flow is strongly dominated by viscous forces, consistent with the behavior of a highly viscous polymer jet.

**S1.3. Calculated Weber number (We):** The Weber number, which characterizes the ratio of inertial forces to surface tension forces, is calculated as [2]:

$$We = \frac{\rho V^2 d}{\sigma} = \frac{(920\,\text{kg/m}^3) \times (0.506\,\text{m/s})^2 \times (2.53 \times 10^{-5}\,\text{m})}{2.5 \times 10^{-2}\,\text{N/m}} \approx 0.238$$

A value of $We \ll 1$ indicates that surface-tension forces overwhelmingly dominate over inertial forces at the scale of the initial jet.

**S1.4. Calculated Ohnesorge number (*Oh*)**: The Ohnesorge number, a key dimensionless parameter governing spinnability, is calculated using the values derived above [3]:

$$Oh = \frac{\sqrt{We}}{Re} = \frac{\sqrt{0.238}}{0.253} \approx 1.93$$

$$Oh = \frac{\mu}{\sqrt{\rho \sigma d}} = \frac{4.654 \times 10^{-2} \text{ Pa·s}}{\sqrt{(920 \text{ kg/m}^3) \times (2.5 \times 10^{-2} \text{ N/m}) \times (2.53 \times 10^{-5} \text{ m})}} \approx 1.93$$

Both methods yield the same results, confirming the internal consistency of the calculations. An Ohnesorge number great than unity (*Oh* > 1) clearly places the solution in the spinnable regime, where viscous forces dominate over inertial and surface tension forces, enabling stable jet elongation and reliable fiber formation.

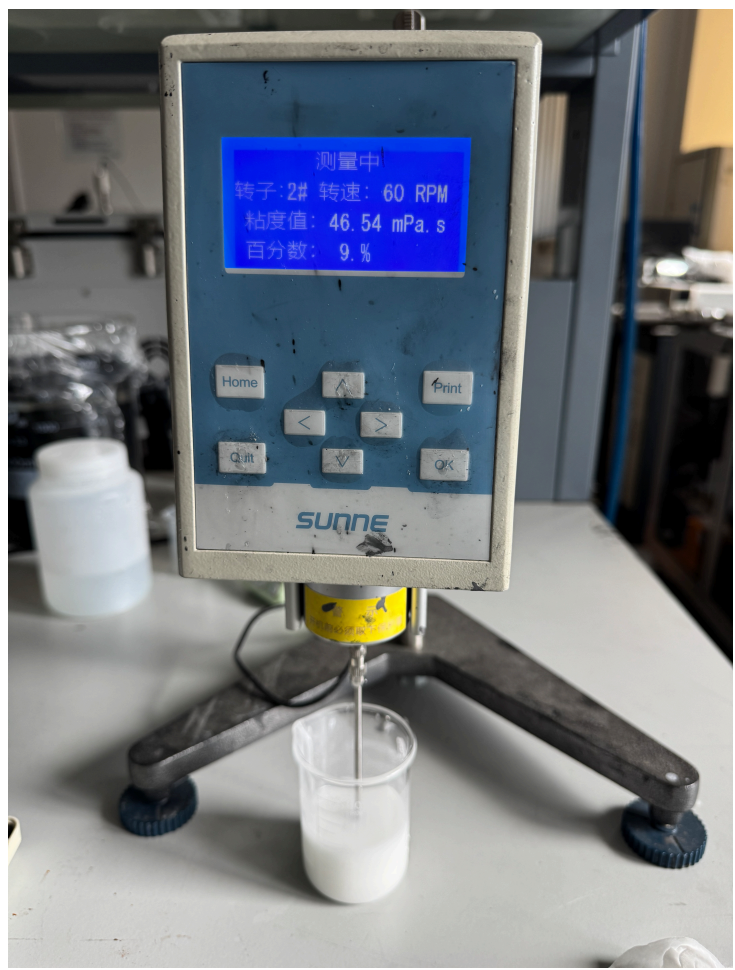

**Figure S1. Viscosity measurement of the SEBS polymer solution.** The dynamic viscosity of the solution was determined to be 46.54 mPa·s using a viscometer (NDJ-8S, Sunne) equipped with a No. 2 spindle operating at 60 rpm. This value was subsequently used to calculate the dimensionless Ohnesorge number (*Oh*).

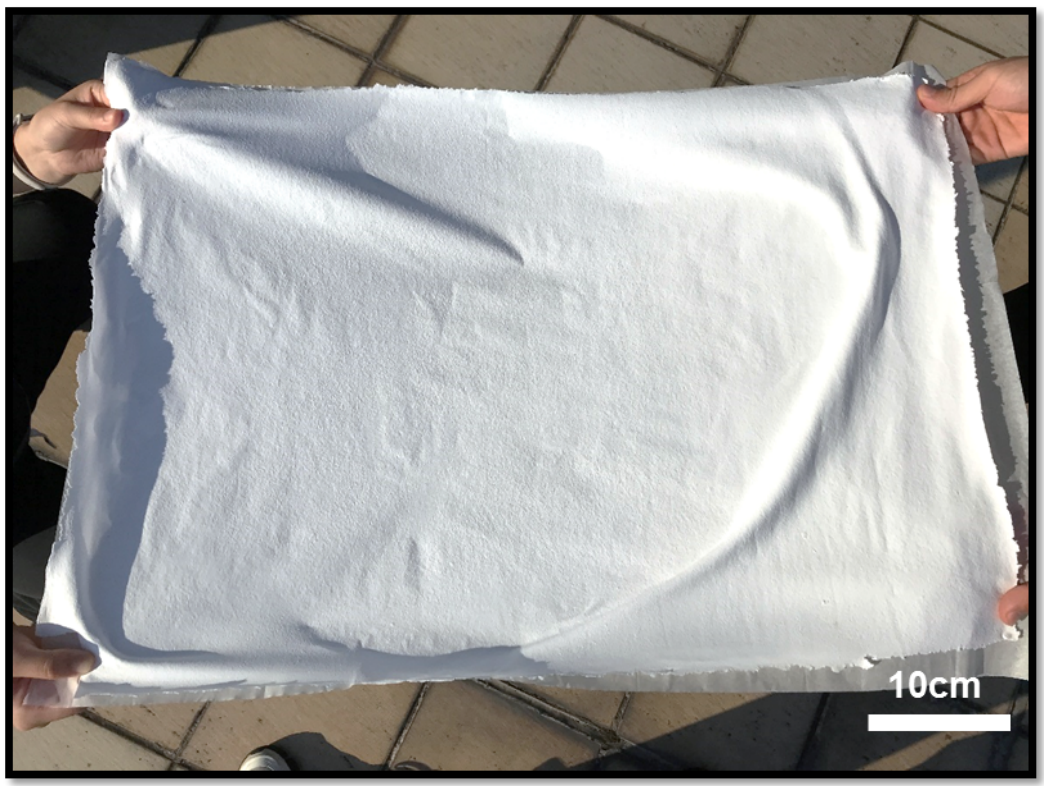

**Figure S2. Photographs of the larger-scale SEBS-Y$_2$O$_3$ composite membrane (50 cm × 70 cm) with a scale bar.** Illustrating the macroscopic uniformity and scalability of the solution blow spinning strategy.

**Note S2. Areal mass, cost analysis, and large-area practicality.**

**Areal mass**：Calculated from experimental data (4 cm×4 cm×0.46 mm membrane, mass 264.67 mg), the areal mass of the single-layer SEBS-$Y_2O_3$ membrane is 165 g/m² (0.165 kg/m²). This lightweight feature avoids excessive substrate load, enabling feasible large-area deployment (e.g., building cladding, outdoor textiles).

**Cost analysis**：Material cost is estimated based on areal mass, raw material market prices (SEBS: 128 CNY/kg; $Y_2O_3$: 456 CNY/kg) and composite mixing ratio. The weighted average total material cost is ~35.3 CNY/m², equivalent to ~5.0 USD/m² (exchange rate: ~7.0 CNY/USD).

**Large-area practicality**：The adopted solution blow spinning method supports high-throughput, large-scale fabrication, which matches the lightweight advantage of the membrane and further verifies its potential for industrial-level deployment.

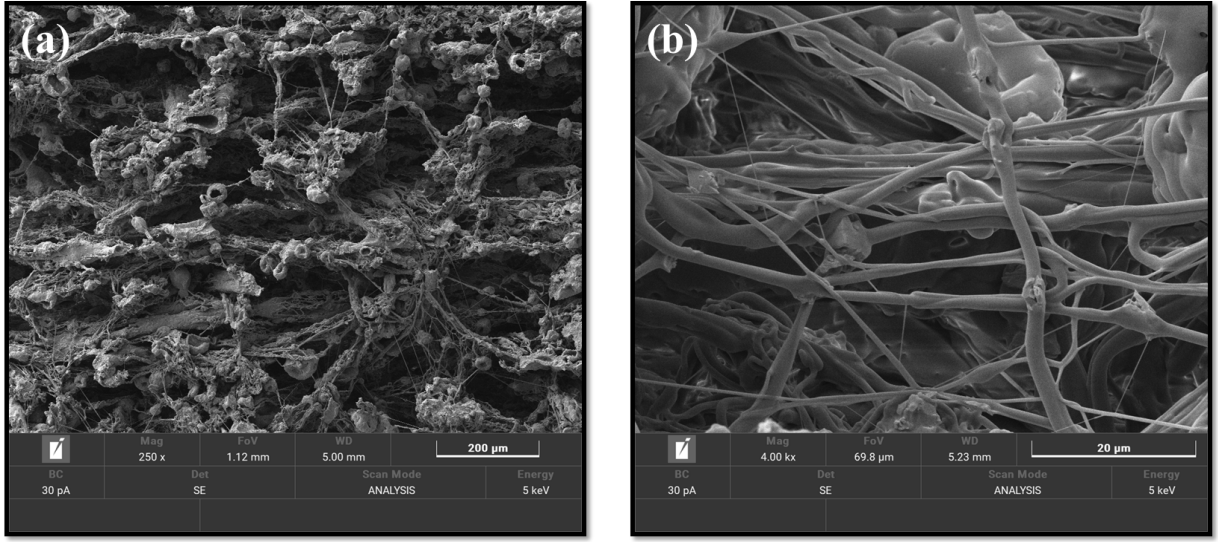

**Figure S3. Cross-sectional SEM image of the single-layer SEBS-Y$_2$O$_3$ fibrous membrane**. Showing a loose and porous fibrous network with uniformly distributed fibers. Such morphology provides structural features relevant to the optical and thermal behavior discussed in the manuscript.

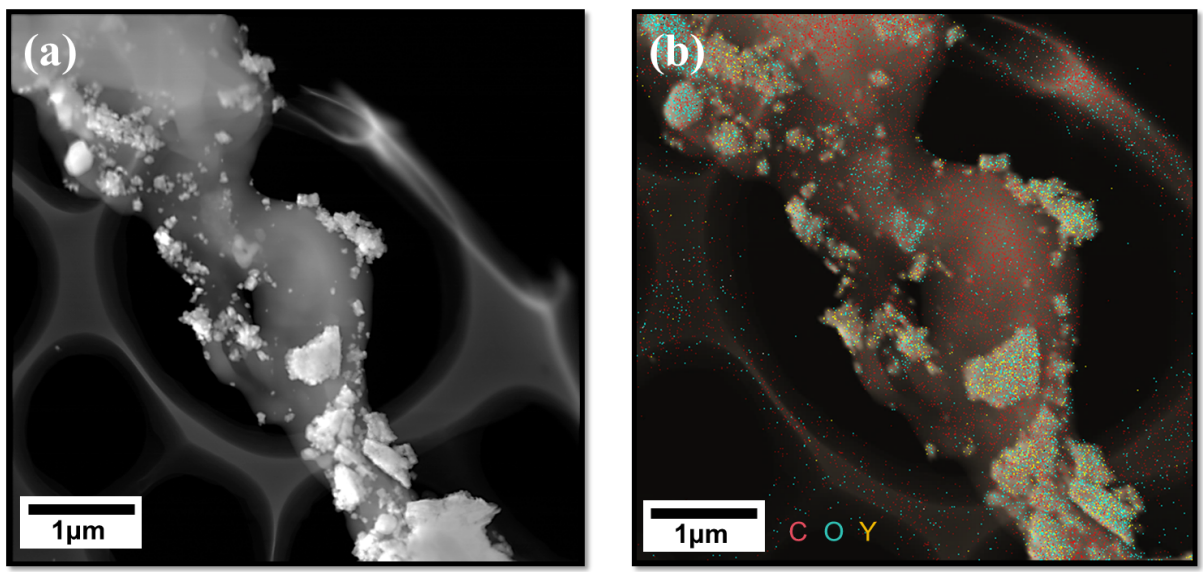

**Figure S4. TEM morphology and EDS elemental mapping of SEBS-Y$_2$O$_3$ fibers.** (a) TEM image of a representative SEBS fiber, in which the dark gray region corresponds to the SEBS polymer matrix and the dispersed bright contrasts are attributed to Y$_2$O$_3$ nanoparticles, without observable large-scale aggregation along the fiber. (b) EDS elemental mapping of the same region, where the red, blue, and yellow signals represent C, O, and Y elements, respectively, confirming the homogeneous dispersion of Y$_2$O$_3$ nanoparticles within the SEBS matrix.

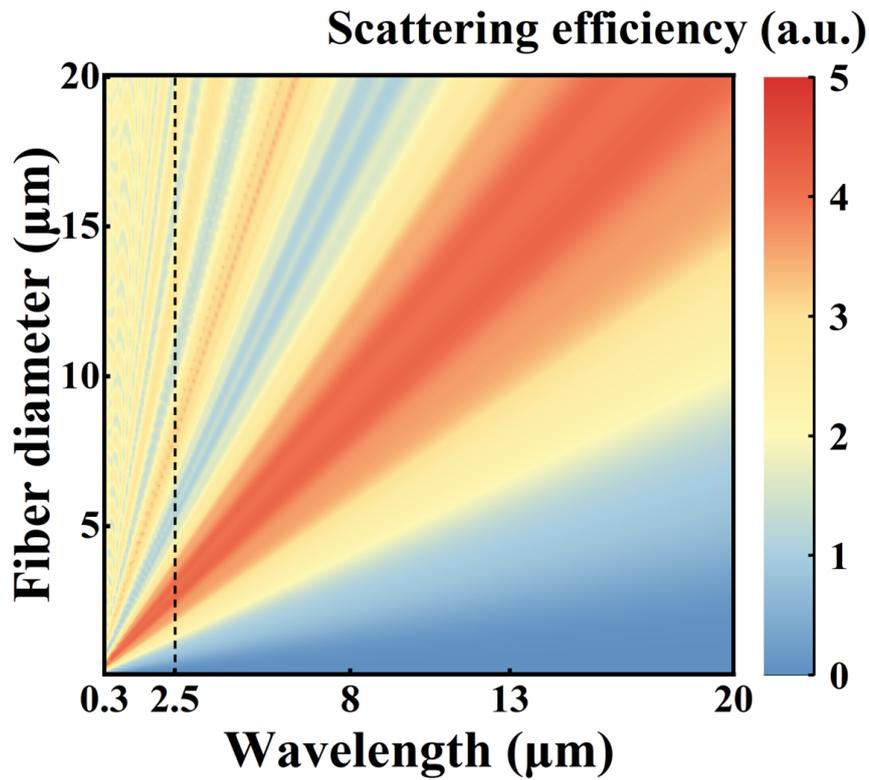

**Figure S5. Simulated scattering efficiency of SEBS fibers as a function of fiber diameters (0–5 μm) and wavelengths (0.3–20 μm).** Red regions indicate high scattering efficiency, whereas blue regions indicate low efficiency. SEBS fibers with a diameter of ~550 nm—consistent with the experimentally measured average diameter of the SEBS-$Y_2O_3$ nanocomposite membrane—exhibit strong scattering across the 0.3–2.5 μm solar spectrum. This effective broadband solar scattering is associated with enhanced solar reflectance relevant to PDRC performance.

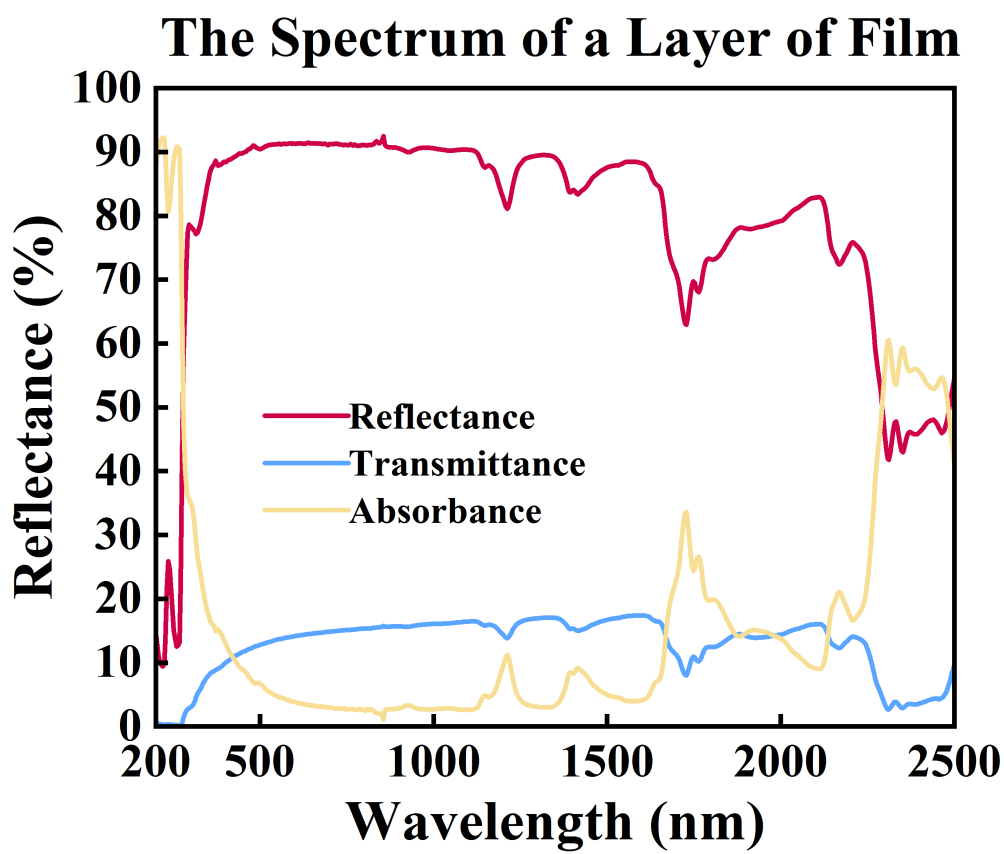

**Figure S6. Optical spectra of the single-layer (0.46 mm) SEBS-Y$_2$O$_3$ membrane in the solar band (300–2500 nm).** Including reflectance (*R*), transmittance (*T*), and absorbance (*A*). The membrane exhibits relatively high transmittance and low absorbance. According to the relationship *R* = 1 – *T* – *A*, the increase in reflectance with increasing thickness is consistent with the reduction in transmittance.

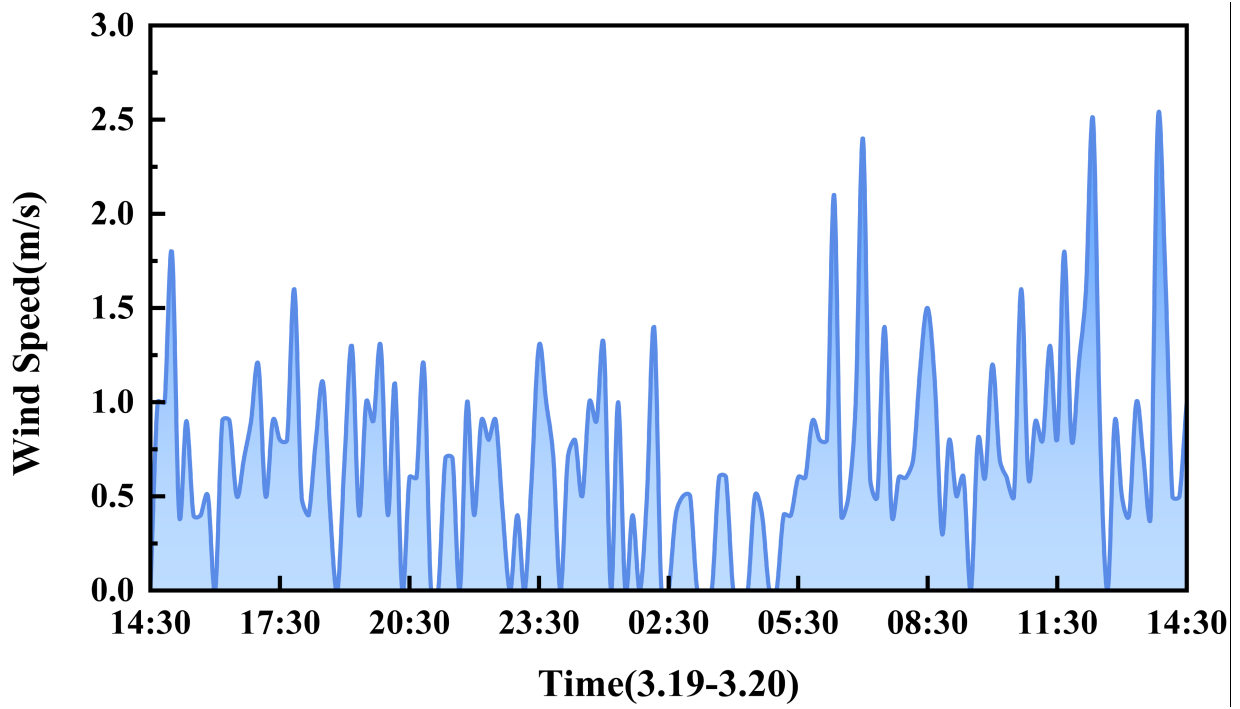

**Figure S7. Diurnal variation of wind speed during the outdoor testing period.** The plot shows the fluctuation of wind speed over a 24-hour cycle, illustrating the natural atmospheric variability that may affect the outdoor cooling performance of the SEBS-$Y_2O_3$ membranes.

### Note S3. Theoretical Calculation of Net Cooling Power

To quantitatively evaluate the PDRC performance of the SEBS-$Y_2O_3$ membranes, we first calculated the net cooling power using the following equation :

$$P_{net}(T) = P_{rad}(T) - P_{atm}(T_{amb}) - P_{sun} - P_{cond+conv}$$

$P_{rad}(T)$ is the radiative power emitted by the membrane, calculated from the measured mid-infrared emissivity spectrum and Planck's blackbody radiation law.

$P_{atm}(T_{amb})$ is the atmospheric radiative power absorbed by the membrane. Herein, the atmospheric emissivity was derived from the atmospheric transmittance data obtained from the Gemini Observatory database, measured at Mauna Kea under clear-sky conditions with air mass 1.5 and a precipitable water vapor (PWV) of 1.0 mm [4]. Considering the more humid subtropical climate of Ningbo, where typical PWV values are substantially higher, the original transmittance was scaled to 55% to approximate local atmospheric conditions. This empirical correction has been widely adopted in PDRC studies to account for enhanced atmospheric absorption in humid environments.

$P_{sun}$ is the solar power absorbed by the membrane, calculated from the measured solar absorptivity spectrum and the standard AM1.5G solar irradiance spectrum (1000 W/m², ASTM G173-03).

$P_{cond+conv} = h_c \cdot (T_{amb} - T)$ is the conduction-convection heat loss, with $h_c$ = 6.9 W/m²·K (experimental combined heat transfer coefficient), $T$ (experimental combined heat transfer coefficient), $T_{amb}$= 301.9K (in-situ ambient temperature).

The steady-state surface temperature $T_s$ satisfies $P_{net}(T_s) = 0$. For the peak daytime period (11:00–13:00, March 20, 2025, solar irradiance ~1000 W/m²), the simulation yields a net cooling power of 18.8 W/m², corresponding to a sub-ambient cooling of 2.0 °C (Fig. S8). Subsequently, a cooling power measurement was conducted over the same period as the outdoor temperature testing.

It should be noted that the reported experimental values of 4.6 °C and 23.1 W/m² are 24-hour weighted averages, which include both daytime and nighttime periods. Nevertheless, for the peak daytime period, the experimental results show an average net cooling power of 18.5 W/m² and a sub-ambient cooling of 2.3 °C, which are in close agreement with the simulation results (18.8 W/m², 2.0 °C). The minor difference (~0.3 °C) is attributed to environmental fluctuations such as ambient temperature variations and wind speed changes during the outdoor test.

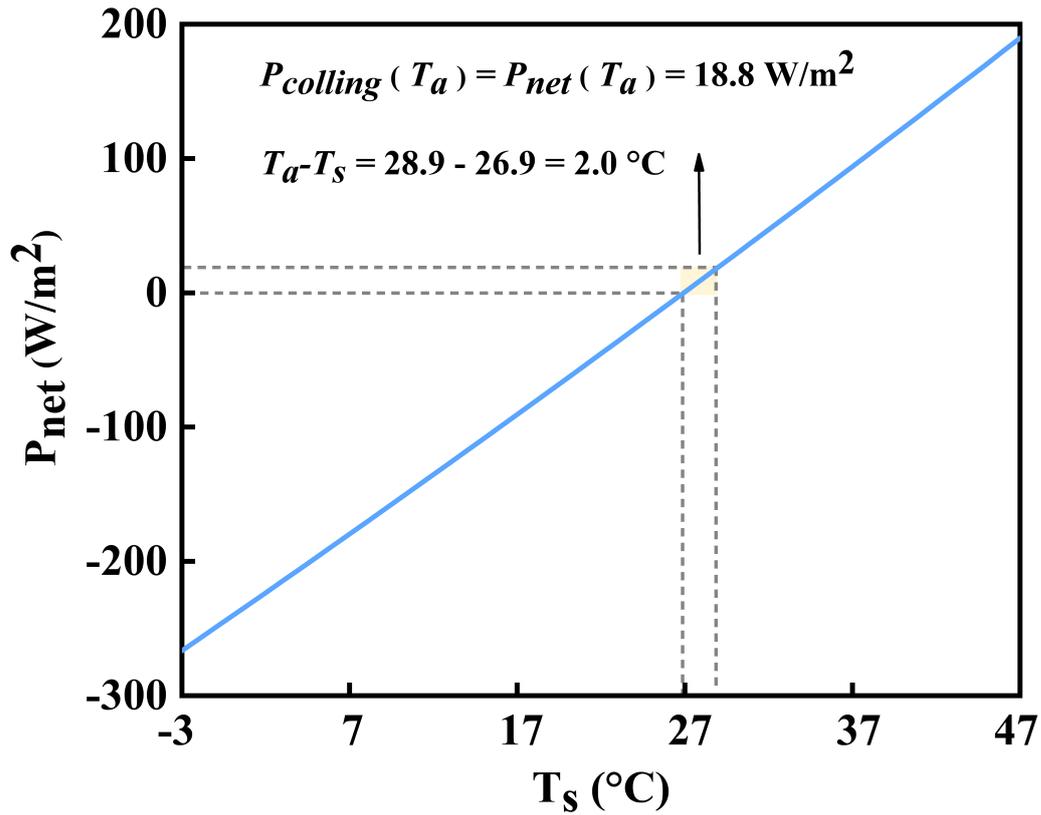

**Figure S8. Simulated correlation between Steady-State Temperature and Net Cooling Power.** Simulated correlation between steady-state surface temperature $T_s$ and net cooling power $P_{net}$ under peak daytime solar irradiance (~1000 W m⁻², 11:00–13:00). The calculated net cooling power of 18.8 W m⁻² corresponds to a sub-ambient cooling of 2.0 °C, which is consistent with the experimental data.

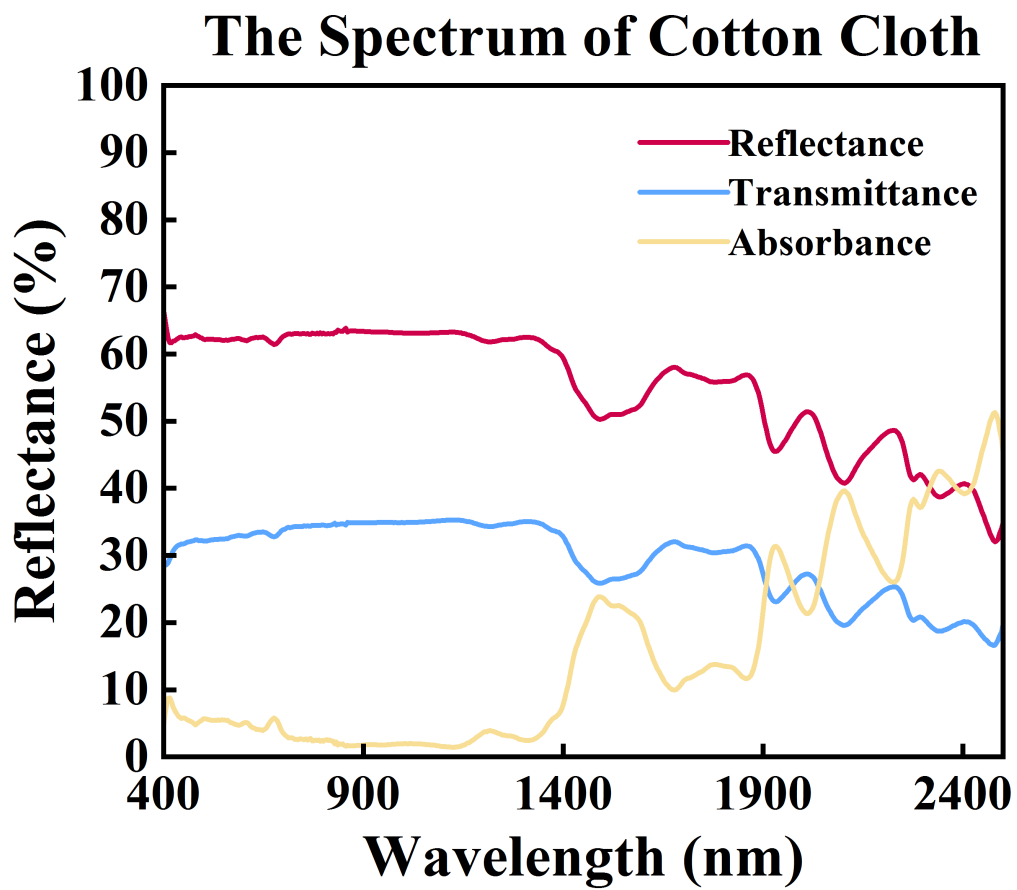

**Figure S9. Solar absorption spectrum of cotton cloth**. The measured spectrum shows relatively low solar absorption, comparable to that of the SEBS-$Y_2O_3$ membrane, supporting its use as a reference material with similar thermal input.

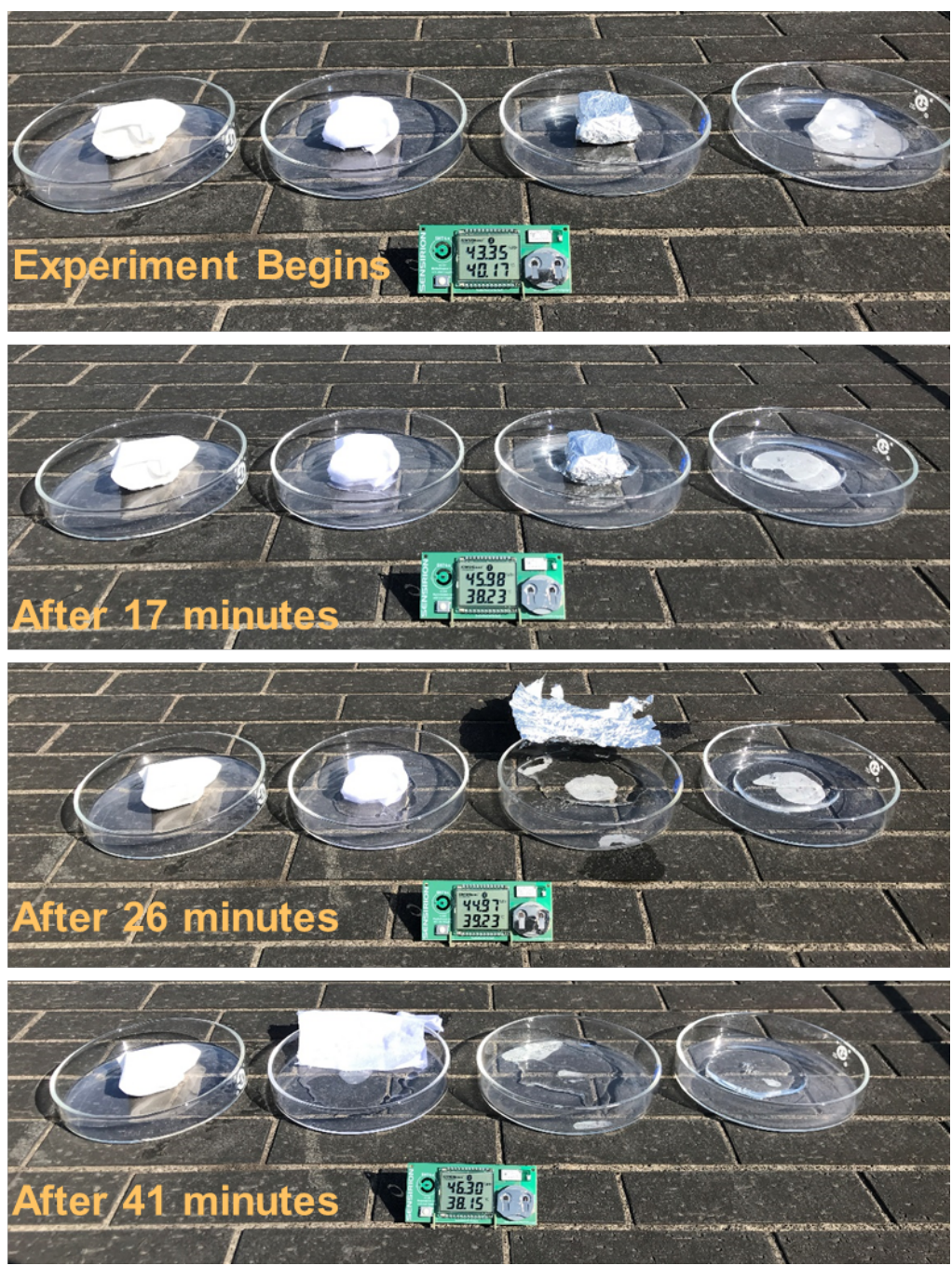

**Figure S10. Outdoor ice-melting inhibition test.** Photographs comparing ice cubes wrapped with different materials (SEBS-$Y_2O_3$ membrane, cotton cloth, aluminum foil) and an unwrapped control under direct sunlight at an average ambient temperature of 38 °C. The SEBS-$Y_2O_3$ membrane demonstrates improved ice-preservation under comparable conditions, retaining a substantial portion of unmelted ice after 41 minutes, whereas all other samples completely melted within the same or shorter time period.

S17

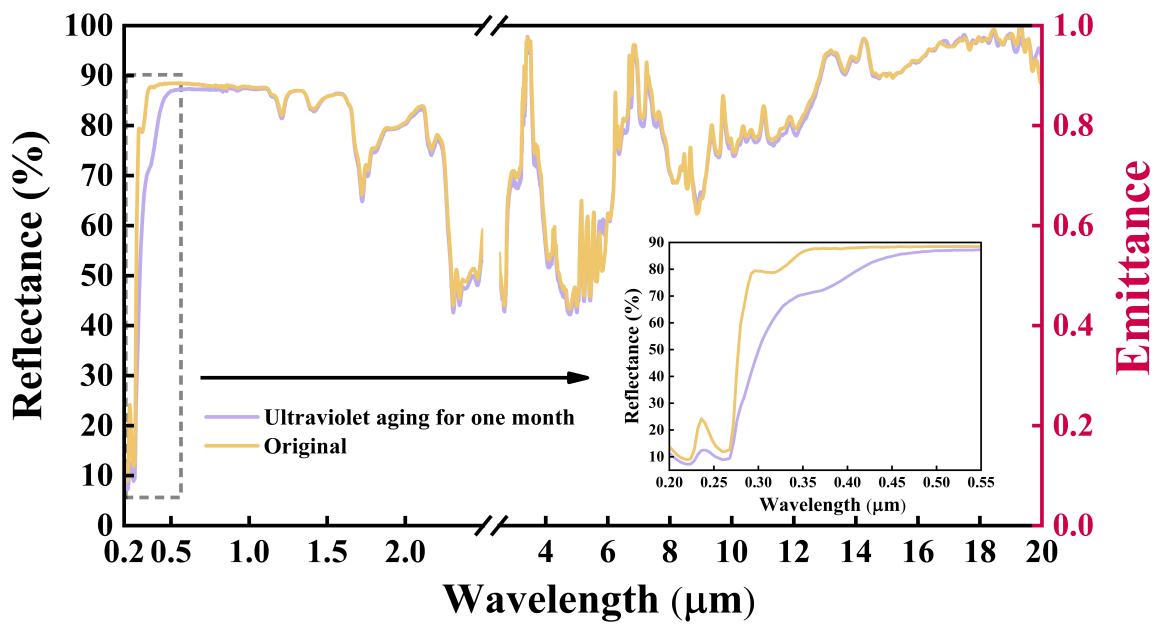

**Figure S11. Optical spectra of the SEBS-Y$_2$O$_3$ membrane before and after one month of UV aging.** Showing an average decrease of ~2% in reflectance within the UV band and nearly unchanged mid-infrared emittance, indicating good stability under UV exposure.

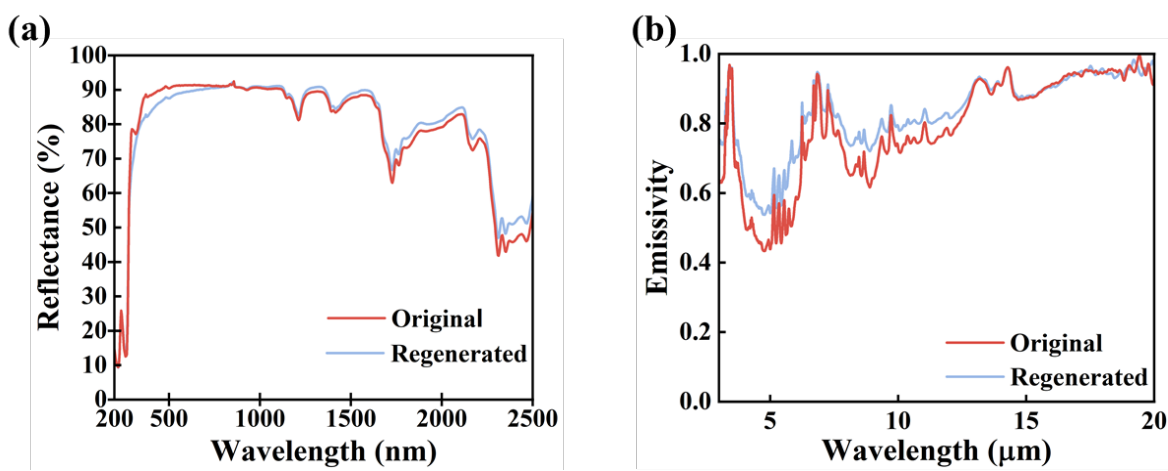

**Figure S12. Comparison of the optical properties between the pristine and the recycled SEBS-Y$_2$O$_3$ membranes.** (a) Solar reflectance and (b) infrared emissivity spectra of the membranes before and after recycling. The recycled membrane largely preserves the high optical performance required for effective PDRC, with variations in solar reflectance and infrared emissivity of only 0.94% and 0.04, respectively.